\documentclass[conference,flushend]{iaria} 
\usepackage{amsmath} 
\usepackage{array}
\usepackage{cleveref}
\usepackage{multicol}
\usepackage{tcolorbox}
\usepackage{pifont}
\usepackage{colortbl}

\title{Practical Acoustic Eavesdropping On Typed Passphrases}

\author{
  \IEEEauthorblockN{Darren Fürst\,\orcidlink{0009-0006-3607-349X}}
  \IEEEauthorblockA{%
    Department of Electrical Engineering, \\
    Media and Computer Science\\
    Ostbayerische Technische Hochschule\\
    Amberg-Weiden, Amberg, Germany\\
    e-mail: \href{mailto:d.fuerst@oth-aw.de}{d.fuerst@oth-aw.de}
  }
  \and
  \IEEEauthorblockN{Andreas Aßmuth\,\orcidlink{0009-0002-2081-2455}}
  \IEEEauthorblockA{%
    Faculty of Computer Science and\\
    Electrical Engineering\\
    Kiel University of Applied Sciences\\
    Kiel, Germany\\
    e-mail: \href{mailto:andreas.assmuth@fh-kiel.de}{andreas.assmuth@fh-kiel.de}
  }
}

\usepackage{ifpdf}
\ifpdf
\pdfoutput=1 
\pdftrue
\fi

\makeatletter
\def\ps@IEEEtitlepagestyle{
\def\@oddfoot{\mycopyrightnotice}
\def\@evenfoot{}
}
\def\mycopyrightnotice{
{\footnotesize
\begin{minipage}{0.8\textwidth}
\centering
Please cite as: Darren Fürst and Andreas Aßmuth ``Practical Acoustic Eavesdropping On Typed Passphrases''
in \emph{Proc of the 16th International Conference on Cloud Computing,
GRIDs, and Virtualization (Cloud Computing 2025), Valencia, Spain, April
2025.}
\end{minipage}
}
}
\makeatother

\begin{document}
\maketitle
\begin{abstract}
Cloud services have become an essential infrastructure for enterprises and individuals. Access to these cloud services is typically governed by Identity and Access Management systems, where user authentication often relies on passwords. While best practices dictate the implementation of multi-factor authentication, it's a reality that many such users remain solely protected by passwords. This reliance on passwords creates a significant vulnerability, as these credentials can be compromised through various means, including side-channel attacks. This paper exploits keyboard acoustic emanations to infer typed natural language passphrases via unsupervised learning, necessitating no previous training data. Whilst this work focuses on short passphrases, it is also applicable to longer messages, such as confidential emails, where the margin for error is much greater, than with passphrases, making the attack even more effective in such a setting. Unlike traditional attacks that require physical access to the target device, acoustic side-channel attacks can be executed within the vicinity, without the user’s knowledge, offering a worthwhile avenue for malicious actors. Our findings replicate and extend previous work, confirming that cross-correlation audio preprocessing outperforms methods like mel-frequency-cepstral coefficients and fast-fourier transforms in keystroke clustering. Moreover, we show that partial passphrase recovery through clustering and a dictionary attack can enable faster than brute-force attacks, further emphasizing the risks posed by this attack vector.
\end{abstract}

\section{Introduction}

As a critical component of modern computing infrastructure, Cloud Services underpin everything from enterprise operations to personal data storage and application access. Securing access to these services is managed through Identity and Access Management (IAM) systems. A fundamental aspect of IAM is user authentication, which, despite the growing adoption of multi-factor authentication, still frequently relies solely on passwords and passphrases. This continued reliance on passwords presents a significant security challenge, as these credentials are vulnerable to a variety of attacks, such as side-channel information leakage. Side-channel attacks aim to infer sensitive information from a system by analyzing unintended emissions, such as power consumption, electromagnetic radiation, or, in the case we explore here, acoustic emanations.

The sounds produced by keyboard typing can reveal valuable information about the typed characters.  While other attacks might require physical proximity to the target device, exploiting acoustic emanations, allow for eavesdropping without user awareness or evidence on the targeted device. This makes acoustic side-channel attacks a realistic and potentially devastating threat to password security. This paper investigates the feasibility of leveraging these keyboard acoustic emanations to infer typed passphrases. We are particularly interested in exploring unsupervised learning techniques, transfering the dictionary demodulation method used by Yang et. al for their WiFi attack~\cite{WirelessTraining-FreeKeystrokeInferenceAttackandDefense}, to the acoustic side-channel. Unsupervised methods offer a more practical approach for real-world attacks as they do not require labeled training data specific to each target user and keyboard. This paper aims to contribute to this understanding by exploring and evaluating methods for acoustic passphrase recovery.

\begin{figure}[!ht]
    \centering
    \includegraphics[width=0.5\textwidth]{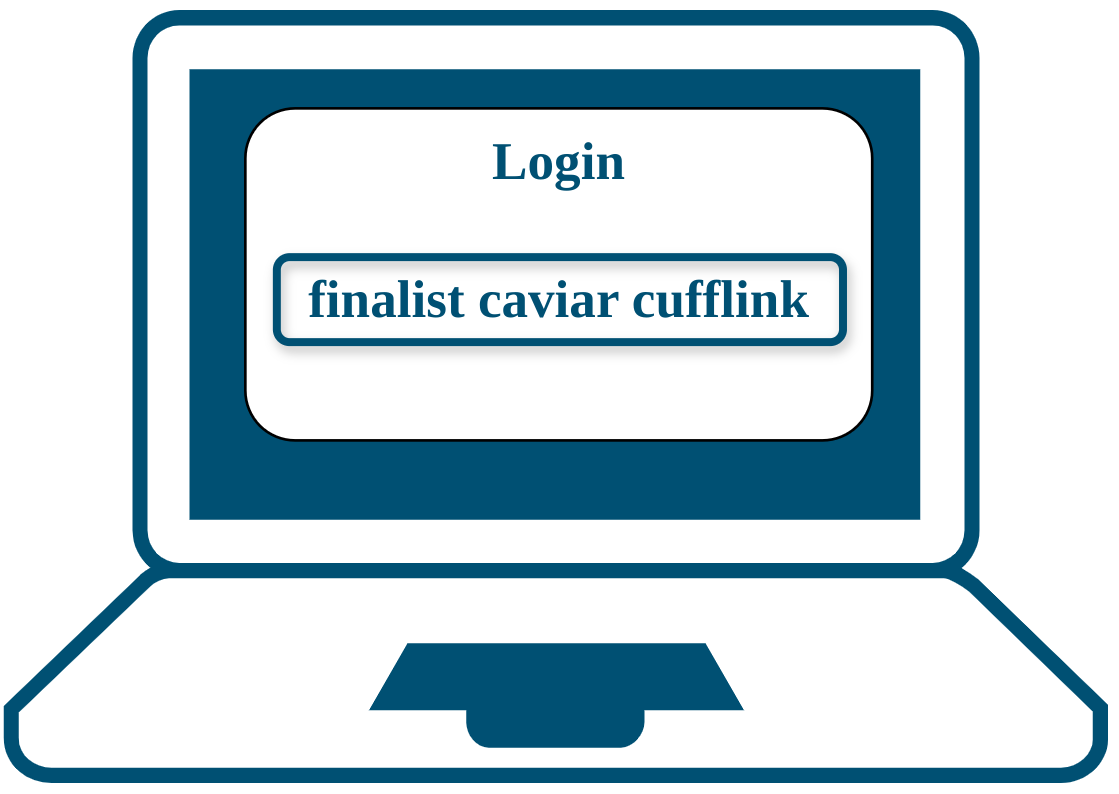}
    \caption{Example of a login screen, where the target types their passphrase to login}
    \label{fig:example_login_screen}
\end{figure}

The rest of the paper is organised as follows: 
\Cref{sec:relatedWork} reviews previous works on side-channel attacks targeting physical user input via keyboards. 
\Cref{sec:typing_mannerisms} discusses typing mannerisms and highlights the challenges posed by various typing styles. 
Next, \Cref{sec:gen_passwds} outlines the methodology behind common password generation, followed by an explanation of the algorithm for passphrase recovery in \Cref{sec:text-recovery}. 
\Cref{sec:experiments} presents the results of hyperparameter tuning, model evaluation, attack performance, and the faster-than-brute-force augmentation technique. 
Finally, \Cref{sec:conclusions} concludes the paper with a summary of the findings and potential future work.

\section{Related Works}\label{sec:relatedWork}

Side-channel attacks have been extensively studied across various modalities, demonstrating the feasibility of inferring sensitive information without directly accessing the target system. In this section, we distinguish between supervised and unsupervised approaches on user input on keyboards.

\subsection{Supervised Approaches}
Supervised methods rely on labeled training data to infer keystrokes or other sensitive information. Whilst demonstrating high accuracy, they are impractical for real-world attacks, as they necessitate collecting labeled data for each target, as well as keyboard.

Asonov and Agrawal~\cite{asonov_keyboard_2004} first demonstrated that keystrokes could be distinguished by analyzing frequency differences, using the Fast Fourier Transform (FFT), to discern between $30$ keys on a keyboard with $79$\,\% accuracy. Subsequent work explored additional features, such as Mel Frequency Cepstral Coefficients (MFCC)~\cite{liu_snooping_2015}\,\cite{pleva_acoustical_2015} and cross-correlation~\cite{Berger_dictionary_acoustic_emantions}\,\cite{halevi2012closer}.

Building on these early studies, recent advances have leveraged deep learning. A deep learning-based approach achieved a classification accuracy of $95\,\%$ on phone-recorded laptop keystrokes and $93\,\%$ on Zoom-recorded audio~\cite{harrison2023practical}. Similarly, Slater et. al. built an end-to-end keystroke segmentation and classification system, achieving a character error rate of $7.41\,\%$ for known typists and $15.41\,\%$ for unknown typists~\cite{10.1145/3359789.3359816}.

Owusu et al. used phone accelerometers to estimate touched screen regions, recovering $59$ out of $99$ six-character passwords~\cite{owusu_accessory_2012}.
Murali et. al. combined acoustic data with motion data from gyrometers to achieve $86\,\%$ accuracy in key recovery using smartphone sensor fusion~\cite{murali2018keyboard}.

By detecting vibrations through accelerometers, Marquardt et al. recovered $80\,\%$ of typed content from a keyboard by placing a mobile device on the same surface~\cite{vibration_inference}. Barisani and Bianco in turn used laser microphones to detect vibrations from laptop screens and utilised a dictionary attack to recover typed words~\cite{barisani2009sniffing}.

Visual-based inference techniques have also been explored. Sabra et al. showed that even subtle upstream movements of the shoulders during typing could be used to recover typed words from video calls~\cite{sabra_zoom_2020}. Moreover, studies have shown that electromagnetic emissions~\cite{vuagnoux2009compromising} and changes in Wi-Fi channel state information~\cite{wikey_2015} can also reveal sensitive keystroke information.

While these supervised methods lay the important groundwork of exploring reliable feature engineering and pre-processing techniques, as well as establishing general feasibility, with many works achieving impressive accuracy in discerning keystrokes, their reliance on labeled data significantly limits their applicability in practical scenarios.

\subsection{Unsupervised Approaches}
Unsupervised approaches, which do not rely on labeled data, present a more promising approach for practical attacks, enabling an attacker to eavesdrop on targets, without prior knowledge and without altering the target's system.

Dictionary-based attacks have been used effectively in recovering typed words from keyboard acoustic emanations, making use of natural language properties. Berger et al. achieved a success rate of $73\,\%$ for $7$ to $13$ character words being in the top $50$ guesses using cross-correlation and a dictionary attack~\cite{Berger_dictionary_acoustic_emantions}. Another method leveraging Time-Difference-of-Arrival (TDoA) measurements from smartphones achieved a $72.2\,\%$ key recognition rate~\cite{zhu2014context}.

Zhuang et. al. used Hidden Markov Models to iteratively generate labels from unlabeled audio recordings, increasing classification accuracy over time. This method recovered up to $96\,\%$ of typed characters from a 10-minute recording~\cite{zhuang_keyboard_2009}. Yang et al. demonstrated an unsupervised Wi-Fi channel-state information attack achieving a $95\,\%$ word recovery ratio after 150 typed words~\cite{WirelessTraining-FreeKeystrokeInferenceAttackandDefense}.

Another attack based on TDoA measurements demonstrated $94\,\%$ keystroke recovery using millimeter-level audio ranging on a single phone~\cite{liu_snooping_2015}.

Whilst some supervised works argue that training data can be recorded via video calls or infected devices, these substantially decrease attack surface and practicality.
In constrast, unsupervised methods, such as employed in this work, provide a feasible manner of eavesdropping via these side-channel mediums, as they do not depend on prior knowledge of the target’s typing style or environment, making them a real threat.

\section{Typing Mannerisms}\label{sec:typing_mannerisms}

Typing ability can affect how a person types a message, with experienced typers typically displaying more consistent typing patterns. This consistency could increase vulnerability to audio-based attacks due to more consistent sounds from their keystrokes. However, their faster typing speed and reduced inter-keystroke pauses might make it harder to distinguish the start and end of keystrokes. In contrast, less experienced typers type more slowly but are likely to have less consistent motions, possibly causing greater variability in sound.

Dhakal et al. analyzed $136$ million keystrokes from $168.000$ volunteers, categorizing typers into eight groups based on metrics such as words per minute and error rates. They found that all groups exhibited at least a $19$\,\% rollover ratio, where multiple keys are pressed consecutively before being released~\cite{dhakal2018observations}. This rollover complicates keystroke segmentation, as it is difficult to determine which press and release belong together. Furthermore, a study of $30$ typers revealed a significant variation in the number of fingers used, with only three using perfect touch typing~\cite{how_we_type}. This highlights the challenges in modeling typing patterns due to the diverse techniques used.

The key challenges are:
\begin{itemize}
    \item Rollover technique complicates keystroke segmentation
    \item Typing error rates vary between typers
    \item Variability in typing styles and proficiency
\end{itemize}

To mitigate these issues, participants were instructed to avoid using rollover patterns for easier segmentation, while recording audio samples. In a real attack, this could be addressed by focusing on initial key presses or using likely press-release combinations. In \Cref{sec:experiments}, both press-only segmentation and press-release segmentation are evaluated for suitability.

\subsection{Selected Features}\label{subsec:selected_features}
Liu et al. used MFCC for K-Means clustering to reduce errors in Time Difference of Arrival measurements~\cite{liu_snooping_2015}. Asonov and Agrawal’s neural network, trained with FFT, achieved $79$\,\% accuracy for the top candidate and $88$\,\% for the top $3$~\cite{asonov_keyboard_2004}. Berger et al.~\cite{Berger_dictionary_acoustic_emantions} and Halevi et. al.~\cite{halevi2012closer} found cross-correlation to outperform FFT and MFCC in keystroke classification, yielding better precision and recall scores. Zhuang et al. showed that using MFCC allowed for correctly classifying more keystrokes than using FFT, their analysis did not include cross-correlation~\cite{zhuang_keyboard_2009}. 

While FFT seems less promising from existing literature than cross-correlation and MFCC, it is included in the evaluation, as it is easily computable.
Thus, the following methods are used alone and in conjunction in the experiments: MFCC, FFT, Cross-Correlation.

\section{Generating Natural Language Passwords}\label{sec:gen_passwds}

This section explains the process of generating natural language passwords used for the attack evaluation.

The UK's National Cyber Security Centre (NCSC) recommends using three random words for constructing passphrases, as adding special characters complicates memorability. They consider passphrases made from three random words to be `strong enough'~\cite{NCSC_password}. Diceware~\cite{Diceware} follows a similar approach, mapping each word to a five-digit number. A word can be looked up by its number, obtained by rolling a dice five times, removing human bias in word selection. The Electronic Frontier Foundation (EFF) has created two wordlists based on this concept, optimised for both memorability and password strength~\cite{EFF_dware}. 

Despite the NCSC's recommendation, humans tend to create weak passwords from a limited set of words~\cite{passwd_sec_a_case_history}. The Yahoo data breach~\cite{Bonneau12} reveals that certain passwords appear far more frequently than others, indicating a strong pattern in human-generated choices. While this chart reflects password frequencies rather than passphrase word frequencies, it suggests that human-generated passphrases may also follow predictable patterns. In contrast, Diceware-generated passphrases benefit from the uniform randomness of the word selection process, making them potentially more secure.

We generate five passphrases each of differing length with $3$ to $8$ words for $30$ passphrases in total from EFF's Long Wordlist to test passphrase recovery. These are shown in \Cref{sec:passphrases}.

\section{Text Recovery}
\label{sec:text-recovery}

The text recovery process can be viewed as breaking a substitution cipher, where cluster indices replace the original alphabetic characters based on keystroke sounds. The final step involves a dictionary attack to map clusters to their correct alphabetic character, producing words. The described method of finding words and demodulating was used in \cite{WirelessTraining-FreeKeystrokeInferenceAttackandDefense} to recover longer typed messages via Wi-Fi channel-state information and is used in this work to recover passphrases formed of $3$ to $8$ words, which would not be possible with $n$-gram statistics or other statistical methods, due to the short message length, via the acoustic side-channel.

\subsection{Finding Words}

Words in natural language are separated by delimiters, typically spaces or hyphens. By leveraging natural language statistics, educated guesses about which cluster represents the delimiter can be made. If the initial guess does not result in a meaningful message, one can iteratively try the next largest cluster \cite{WirelessTraining-FreeKeystrokeInferenceAttackandDefense}. In a passphrase with $n$ words, the delimiter appears $n-1$ times and is thus likely one of the larger clusters.

\subsection{Inter-Element Relationship Matrix}

To identify word candidates, we use features such as word length, letter frequencies, and same-letter positions. An inter-element relationship matrix \cite{WirelessTraining-FreeKeystrokeInferenceAttackandDefense} is constructed, where letters are compared and marked with 1 for identical letters and 0 for differing ones. This results in a symmetrical matrix, which describes each word or concatenation of words by length and frequencies and positions of same letters.

\begin{figure}[!ht]
    \centering
    \begin{minipage}{0.24\textwidth} 
        \centering
        \resizebox{\textwidth}{!}{
        \begin{tikzpicture}
            \definecolor{customRed}{RGB}{255,102,102}
            \definecolor{customBlue}{RGB}{102,178,255}
            \definecolor{customGreen}{RGB}{144,238,144}
            \definecolor{customViolet}{RGB}{238,130,238}

            \node at (0,0) {
                $
                \begin{array}{c|ccccc}
                & l & e & v & e & l \\
                \hline
                l & 1 & \cellcolor{customRed}0 & \cellcolor{customRed}0 & \cellcolor{customRed}0 & \cellcolor{customRed}1 \\
                e & 0 & 1 & \cellcolor{customBlue}0 & \cellcolor{customBlue}1 & \cellcolor{customBlue}0 \\
                v & 0 & 0 & 1 & \cellcolor{customGreen}0 & \cellcolor{customGreen}0 \\
                e & 0 & 1 & 0 & 1 & \cellcolor{customViolet}0 \\
                l & 1 & 0 & 0 & 0 & 1 \\
                \end{array}
                $
            };
        \end{tikzpicture}
        }
    \end{minipage}
    \hfill
    \begin{minipage}{0.24\textwidth} 
        \centering
        \resizebox{\textwidth}{!}{
        \begin{tikzpicture}
            \definecolor{customRed}{RGB}{255,102,102}
            \definecolor{customBlue}{RGB}{102,178,255}
            \definecolor{customGreen}{RGB}{144,238,144}
            \definecolor{customViolet}{RGB}{238,130,238}

            \node at (0,0) {
                $
                \begin{array}{c|ccccc}
                & r & a & d & a & r \\
                \hline
                r & 1 & \cellcolor{customRed}0 & \cellcolor{customRed}0 & \cellcolor{customRed}0 & \cellcolor{customRed}1 \\
                a & 0 & 1 & \cellcolor{customBlue}0 & \cellcolor{customBlue}1 & \cellcolor{customBlue}0 \\
                d & 0 & 0 & 1 & \cellcolor{customGreen}0 & \cellcolor{customGreen}0 \\
                a & 0 & 1 & 0 & 1 & \cellcolor{customViolet}0 \\
                r & 1 & 0 & 0 & 0 & 1 \\
                \end{array}
                $
            };
        \end{tikzpicture}
        }
    \end{minipage}
    \caption{Example of two words, with the same inter-element relationship matrix, although their letters differ. The coloring is added to enable quick comparison of the symmetrical matrix.}
\end{figure}

\subsection{Joint Demodulation}

The candidate selection and dismissal process is based on the Joint Demodulation method from Yang et al.~\cite{WirelessTraining-FreeKeystrokeInferenceAttackandDefense}. This involves concatenating candidate words from a dictionary and comparing their inter-element relationship matrix with the matrix of the audio cluster. Concatenations resulting in a different inter-element relationship matrix are discarded as potential passphrases. If no words are found for a concatenation, the last appended word is skipped and added to the undemodulated set \cite{WirelessTraining-FreeKeystrokeInferenceAttackandDefense}, where it is later resubstituted with the letter-mappings found by demodulating the concatenation of the remaining words.

\section{Experiments}\label{sec:experiments}

The experiments were conducted using the Diceware Long Wordlist~\cite{Diceware} as a dictionary. 

\subsection{Hyperparameter Search}\label{sec:eval_diff_models}

To identify the most suitable clustering model, a hyperparameter search was conducted for two model types, with $n$ being the amount of configurations tested: K-Means ($n = 2049$) and Cross-Correlation ($n = 2045$). The Cross-Correlation type computes the correlation of keystroke segements based on the recorded raw audio, MFCC or FFT transformation of the audio, before clustering with K-Means, while K-Means uses the feaature vectors gained from applying MFCC or FFT, directly. This naming distinction is used to be able to talk about and distinguish these model types. To avoid overfitting of the hyperparameters to the whole dataset, skewing recovery results, $20$ samples from the participants were picked at random and used in the search, spanning $3$ to $5$ samples per participant.

The keystroke span `PR' uses both press and release events, while `P' uses only the key press. The window size for these events was manually set.

An optional convolutional smoothing step was applied, with window sizes included in the hyperparameter search.

\begin{table}[!ht]
    \centering
    \caption{Hyperparameters for K-Means and Cross-Correlation-based Models.}
    \resizebox{0.5\textwidth}{!}{
    \rowcolors{2}{gray!15}{white}
    \begin{tabular}{>{\raggedright\arraybackslash}p{2.4cm}>{\centering\arraybackslash}p{3.2cm}>{\centering\arraybackslash}p{3.2cm}}
    \rowcolor{gray!30} 
    \textbf{Hyperparameter} & \textbf{K-Means} & \textbf{Cross-Correlation} \\ 
    \hline
    Feature & FFT, MFCC, FFT+MFCC & Raw Audio, FFT, MFCC \\ 
    Smoothing & \multicolumn{2}{c}{true, false} \\ 
    Smoothing Window & \multicolumn{2}{c}{5 to 300} \\ 
    Scaling & \multicolumn{2}{c}{true, false} \\ 
    PCA & \multicolumn{2}{c}{true, false} \\ 
    PCA Components & 1 to 20 & 1 to 12 \\ 
    Keystroke Span & \multicolumn{2}{c}{P, PR} \\ 
    \end{tabular}
    }
    \label{tab:combined_hyperparams}
\end{table}

The best models by median score of each type are shown in \Cref{tab:best_models}. Cross-Correlation, using raw audio, outperformed K-Means, which was most effective using MFCC and Principal Component Analysis (PCA).

\begin{table}[!ht]
    \centering
    \caption{Best Model Scores and Their Hyperparameters.}
    \resizebox{0.5\textwidth}{!}{
    \rowcolors{2}{gray!15}{white}
    \begin{tabular}{>{\raggedright\arraybackslash}p{3.2cm}>{\centering\arraybackslash}p{2.7cm}>{\centering\arraybackslash}p{2.7cm}}
    \rowcolor{gray!30}
    \textbf{Hyperparameter} & \textbf{K-Means} & \textbf{Cross-Correlation}  \\ 
    \hline
    \textbf{Feature} & MFCC & Raw  \\ 
    \textbf{MFCC Components} & 180 & \\ 
    \textbf{PCA} & True & False \\ 
    \textbf{PCA Components} & 1 &  \\ 
    \textbf{Smoothing} & False & False  \\ 
    \textbf{Scaling} & True & False \\ 
    \textbf{Keystroke span} & PR & P \\ 
    \textbf{Median Score} & 90.27 & 93.12  \\ 
    \textbf{Mean Score} & 88.95 & 93.21  \\ 
    \textbf{Max Score} & 91.77 & 98.91 \\ 
    \textbf{Min Score} & 83.07 & 85.44  \\ 
    \end{tabular}
    }
    \label{tab:best_models}
    \normalsize
\end{table}

Despite previous works clearly favouring MFCC, FFT was competitive in K-Means models, showing that FFT can achieve comparable performance under the right hyperparameter configurations. The top three models per type, with their respective audio feature processing are summarised in \Cref{tab:top_models}. This shows that with a more extensive hyperparameter search the top models are very close to the same scores.

\begin{table}[ht]
    \centering
    \caption{Top 3 Models per Type and Their Scores.}
    \resizebox{0.5\textwidth}{!}{ 
    \large
    \rowcolors{2}{gray!15}{white}
    \begin{tabular}{>{\raggedright\arraybackslash}p{3.2cm}>{\raggedright\arraybackslash}p{1.5cm}>{\centering\arraybackslash}p{1.8cm}>{\centering\arraybackslash}p{1.8cm}>{\centering\arraybackslash}p{1.8cm}>{\centering\arraybackslash}p{1.8cm}} 
    \rowcolor{gray!30} 
    \textbf{Model Type} & \textbf{Feature} & \textbf{Median} & \textbf{Mean} & \textbf{Max} & \textbf{Min} \\ 
    \hline
    K-Means              & MFCC           & 90.27 & 88.95 & 91.77 & 83.07 \\ 
    K-Means              & FFT            & 89.96 & 88.41 & 92.47 & 79.84 \\ 
    K-Means              & FFT            & 89.95 & 88.62 & 92.90 & 79.89 \\ 
    Cross-Correlation   & Raw            & 93.12 & 93.21 & 98.91 & 85.44 \\ 
    Cross-Correlation   & Raw            & 92.85 & 93.18 & 98.27 & 87.29 \\ 
    Cross-Correlation   & Raw            & 92.85 & 93.52 & 99.13 & 88.36 \\ 
    \end{tabular}
    }
    \label{tab:top_models}
    \normalsize
\end{table}

Cross-Correlation models showed superior performance, especially with raw audio features, while K-Means models using MFCC or FFT performed similarly. This suggests that hyperparameter choices, particularly feature extraction and preprocessing, significantly impact clustering effectiveness for acoustic eavesdropping.

\subsection{Recovering Passphrase Recordings}\label{sec:recov_passphrases}

The best general model, which is of the Cross-Correlation type, from the hyperparameter search on the subset of participant samples was used to cluster a total of $223$ samples. The hyperparameter search and selection of the best model is explained in \Cref{sec:eval_diff_models}. The top model configuration per type with hyperparameters and scores is shown in \Cref{tab:best_models}. As the recording process was conducted via a custom built website to keep the recording manner similar between participant's, some participants' microphones removed keystroke sounds for the samples due to in-built noise reduction features. Such samples were discarded after listening.
For the experiments in-built laptop microphones were used, as this made recording simply feasible via the custom built website. However, in a real-world scenario an attacker would most likely plant their own microphone, as having access to the target machine's microphone would mean the machine has already been compromised, removing such challenges, as built-in noise reduction. Furthermore, an attacker may use more high-end hardware, whereas this study aims to show feasability with even low-cost hardware, such as the built-in microphones used.
The usable samples per participant are shown in \Cref{tab:samples_per_participant}.

\begin{table}[!ht]
    \centering
    \caption{Samples per participant}
    \resizebox{0.18\textwidth}{!}{  
    \small
    \rowcolors{2}{gray!15}{white}
    \begin{tabular}{>{\centering\arraybackslash}p{1.3cm}>{\centering\arraybackslash}p{1.5cm}} 
    \rowcolor{gray!30} \textbf{Participant} & \textbf{Passphrases} \\ 
    \hline
    1 & 30 \\ 
    2 & 30 \\ 
    3 & 16 \\ 
    4 & 30 \\ 
    5 & 19 \\ 
    6 & 27 \\ 
    7 & 22 \\ 
    8 & 30 \\ 
    9 & 19 \\ 
    \end{tabular}
    }
    \label{tab:samples_per_participant}
\end{table}

In a real-world attack, words from the undemodulated set \cite{WirelessTraining-FreeKeystrokeInferenceAttackandDefense}, would have to be checked against a large dictionary to find the correct candidate word, as the words from the undemodulated set likely contain a cluster assignment error, which can be resolved by checking against known English words.
To simulate such a dictionary correction, the following Hamming distance per word length was deemed as corrected, by such a dictionary:

\begin{equation*}
    \text{Hamming~Allowance}(w) =
        \begin{cases}
            0 & \text{if } \#w \leq 2 \\
            1 & \text{if }~3 \leq \#w \leq 4 \\
            2 & \text{if }~5 \leq \#w \leq 6 \\
            3 & \text{if }~7 \leq \#w \leq 9 \\
        \end{cases}
\end{equation*}

\begin{figure}[!ht]
    \centering
    \includegraphics[width=0.5\textwidth]{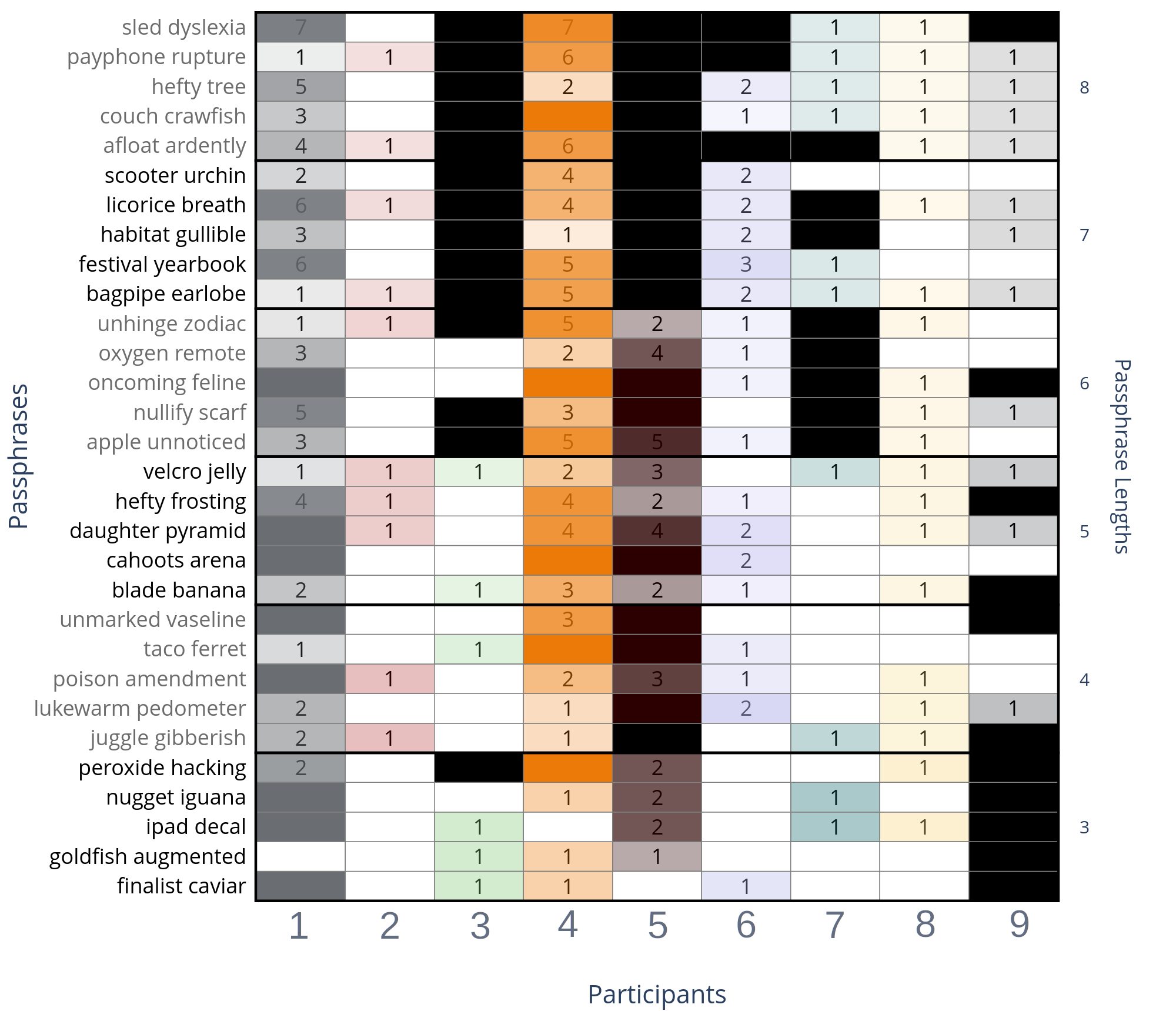}
    \caption{Recovery results using ten clusters.}
    \label{fig:recovery_results_rectangles_ten_clusterings}
\end{figure}

\Cref{fig:recovery_results_rectangles_ten_clusterings} shows the recovery results, where full recoveries are marked with fully coloured rectangles, partial recoveries with partial colouring along with the amount of recovered words, and unrecoverable passphrases with colourless rectangles. Black rectangles represent unusable or samples not recorded by participants. The first two words of each passphrase are shown. The bottom $5$ passphrases are $3$ words long up to the top $5$ passphrases having a length of $8$ words. The full passphrase list is listed in the \Cref{sec:passphrases}.

\begin{table}[!ht]
    \centering
    \caption{Hardware used by participants.}
    \resizebox{0.5\textwidth}{!}{  
    \normalsize
    \rowcolors{2}{gray!15}{white}
    \begin{tabular}{>{\centering\arraybackslash}p{1.5cm}>{\centering\arraybackslash}p{3.2cm}>{\centering\arraybackslash}p{3.7cm}>{\centering\arraybackslash}p{1.7cm}} 
    \rowcolor{gray!30}
    \textbf{Participant} & \textbf{Keyboard Model} & \textbf{Microphone Model} & \textbf{Mechanical} \\ 
    \hline
    1  & Tecurs                        & IdeaPad 5 Pro 14ACN6      & \ding{51} \\ 
    2  & Laptop                        & Laptop Webcam             & \ding{55} \\ 
    3  & Apple Magic (2014) & iMac 2014                 & \ding{55} \\ 
    4  & HIGROUND Base 65              & Auna CM 900B              & \ding{51} \\ 
    5  & Keychron K8 Pro               & MacBook Air M1            & \ding{51} \\ 
    6  & Redragon                      & Macbook Pro 14            & \ding{51} \\ 
    7  & Cherry                        & Laptop                    & \ding{51} \\ 
    8  & Corsair K55 Gaming            & Lenovo ThinkPad T14s      & \ding{55} \\ 
    9  & Cherry                        & DELL Notebook             & \ding{51} \\ 
    \end{tabular}
    }
    \label{tab:participants_hardware}
\end{table}

Mechanical keyboards were more susceptible, likely due to their louder and more distinct sound profiles, although typing styles, microphone quality, and background noise likely also contributed. 

Recovery rates improved using multiple sets of clusters. By applying $10$ sets of clusters over a single cluster attempt by the model, shown in \Cref{app:single_cluster_recovery}, full recovery increased from $7$ to $19$ passphrases, primarily from the same highly susceptible participants. This also boosted partial recoveries. For example, a sample for participant `5` seeing improvements from $3$ to $6$ recovered words (\autoref{fig:recovery_results_rectangles_ten_clusterings}).

In conclusion, a single clustering set achieved partial recovery for all participants, while $10$ sets improved full recoveries to $19$ and enhanced partial recovery success. A further plot showing the recovery increase for different amount of cluster sets can be seen in \Cref{app:single_cluster_recovery}.

\subsection{Brute-Forcing Combinations of Different Recoveries}\label{sec:brute_force_recov}

An attacker can use the words found by partial recoveries in a brute-force attack by forming the product of these words.

\begin{figure}[!ht]
    \centering
    \includegraphics[width=0.5\textwidth]{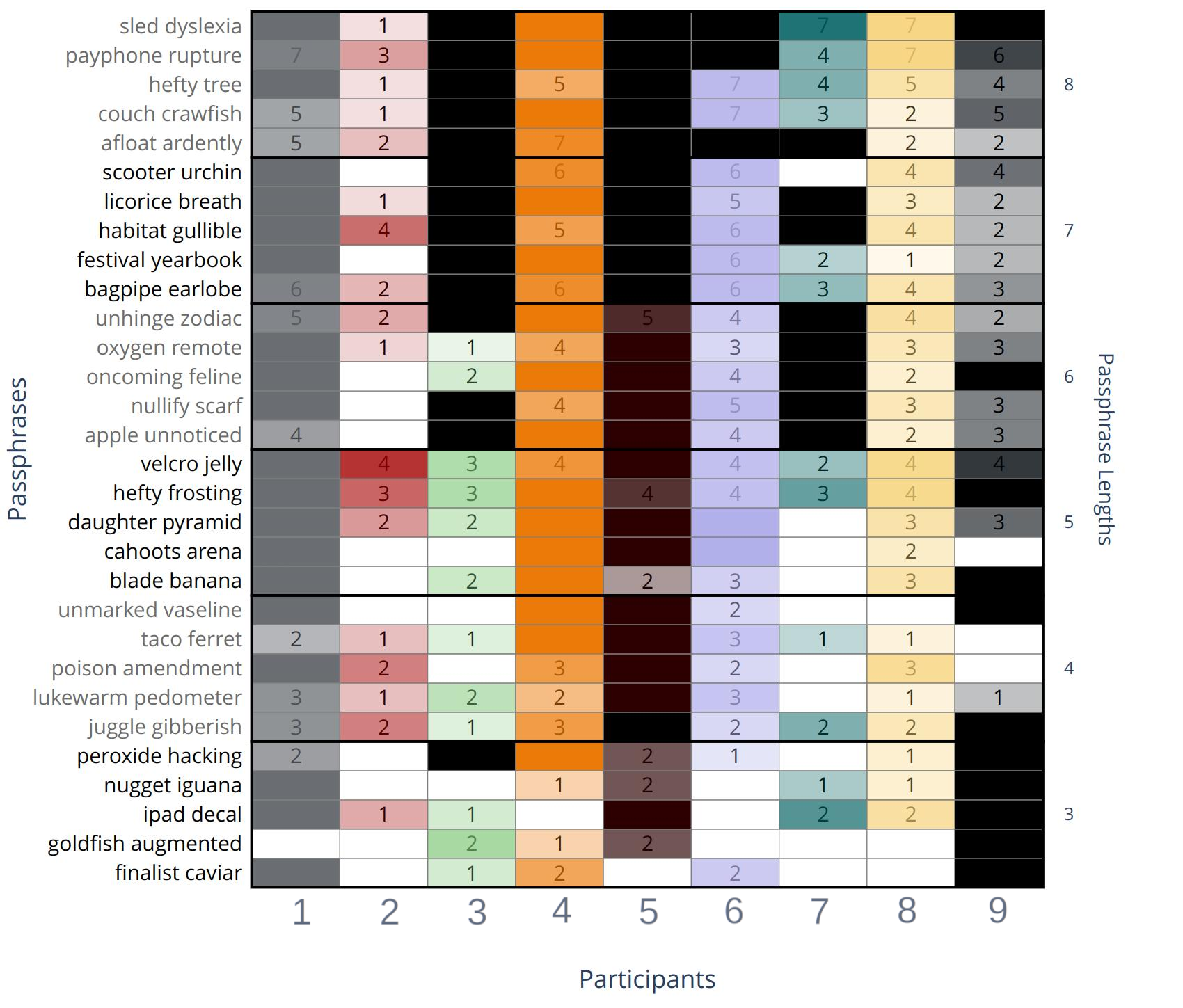}
    \caption{Recoveries brute-forcing combinations of partial recoveries from ten clusters.}
    \label{fig:combinations_recovery_matrix}
\end{figure}

\begin{figure}[!ht]
    \centering
    \includegraphics[width=0.5\textwidth]{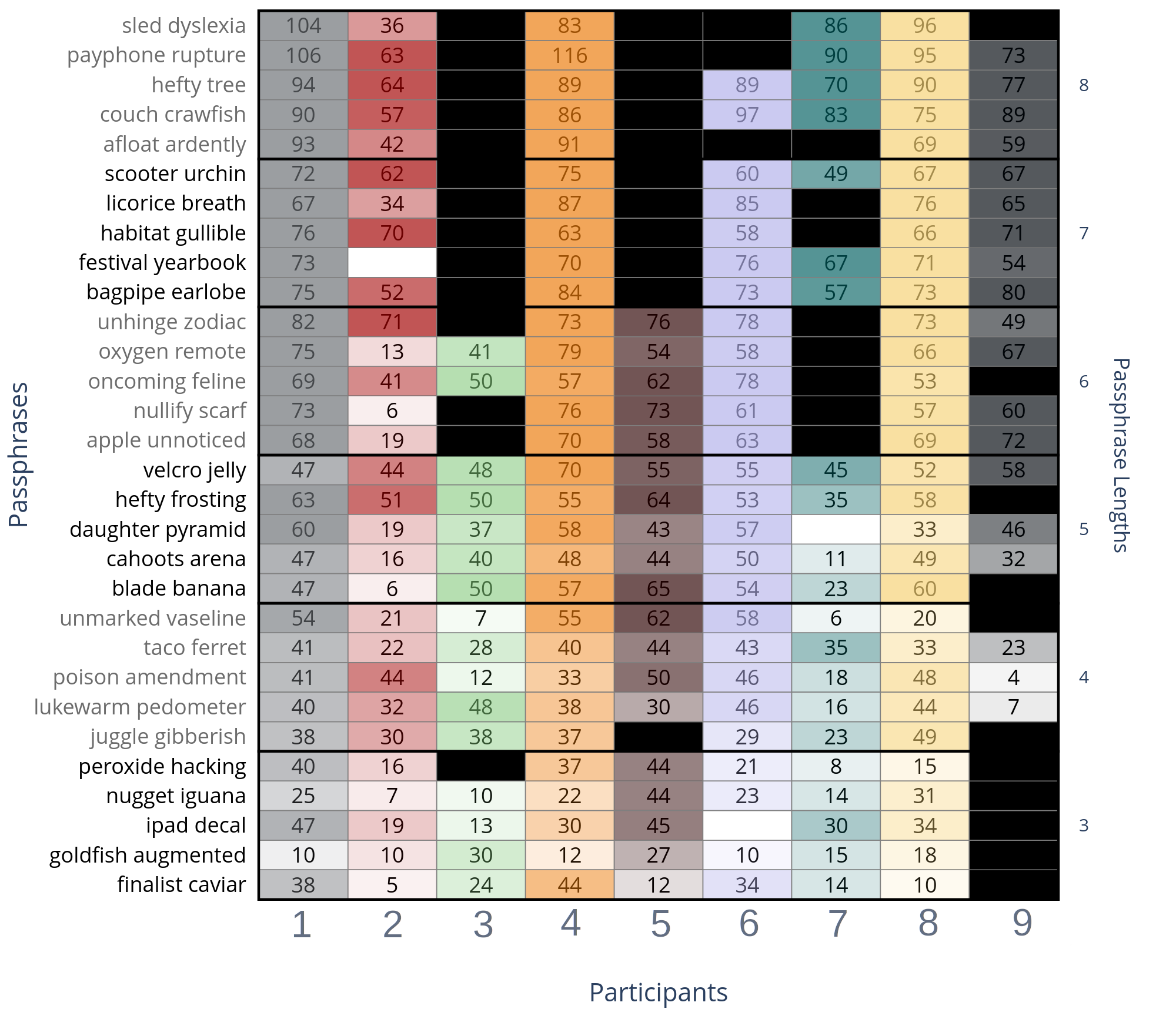}
    \caption{Amount of combinations of demodulated words from ten cluster results. The table shows the exponents to the base of 2.}
    \label{fig:combinations_demod_word_base_2}
\end{figure}

\Cref{fig:combinations_recovery_matrix} shows the recovery results for brute-forcing combinations of words from partial recoveries, by adding each found word at each index to a set and forming the combinations. The number of combinations needed for a brute-force attack is illustrated in \Cref{fig:combinations_demod_word_base_2}, with exponents representing the possible combinations. For example, participant `1' has $2^{38}$ possible combinations from their partial recoveries for the first passphrase 'finalist caviar cufflink` (bottom left).

\begin{figure}[!ht]
    \centering
    \includegraphics[width=0.5\textwidth]{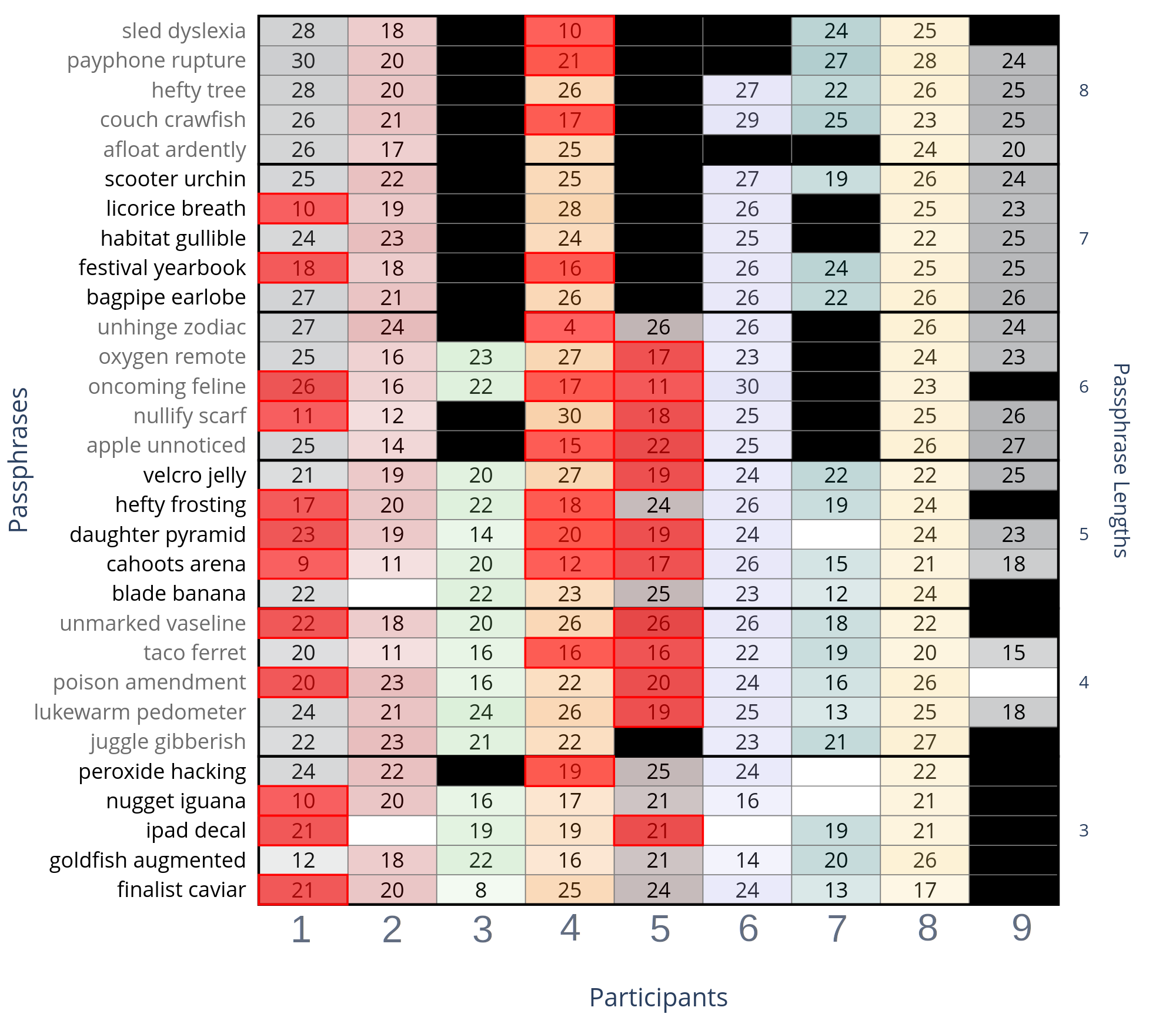}
    \caption{Brute-Force attempts needed, when starting with most likely candidates from ten cluster results. The table shows the exponents to the base of 2. Red marked cells are full recoveries.}
    \label{fig:brute_force_combinations_demod_word_base_2}
\end{figure}

However, brute-forcing all combinations naively this way disregards the position of the found words and is still computationally expensive. An alternative approach, starting with the most likely candidate and adding missing words, reduces the number of required combinations (\autoref{fig:brute_force_combinations_demod_word_base_2}). This method, though more efficient, can still fail to fully recover the passphrase, as shown by the red rectangles marking complete successful recoveries. Evidently, there are less full recoveries than in \Cref{fig:combinations_recovery_matrix}, but not all full recoveries in \Cref{fig:combinations_recovery_matrix}, would be computable by even the strongest adversaries, as shown in \Cref{fig:combinations_demod_word_base_2}, with multiple recoveries needing more than $2^{80}$ steps.

In conclusion, starting with partial recoveries and narrowing down candidate words reduces the computational cost of brute-forcing below the border of computational feasibility in terms of complexity theory. This method can be further optimised by leveraging multiple clustering sets to account for errors in the clusters.

\vspace{1cm}
\section{Conclusion and Future Work}\label{sec:conclusions}

This study demonstrates that attackers can effectively recover passphrases from audio data, even without direct access to typed text, making the attack a potential non-intrusive and passive part of an attack chain, depending on whether the target has multi-factor authentication in place or not. 

The results confirm previous findings~\cite{Berger_dictionary_acoustic_emantions}\,\cite{halevi2012closer}, showing that cross-correlation outperforms MFCC and FFT for keystroke clustering. However, it also showed that MFCC and FFT remain competitive under certain hyperparameters, suggesting the need for further parameter and model exploration. The hyperparameter search was conducted across  $20$ audio recordings from nine participants.
Additionally, the dictionary attack by Yang et al.~\cite{WirelessTraining-FreeKeystrokeInferenceAttackandDefense} was adapted to the acoustic side-channel and
an attack exploiting partial passphrase recoveries with significant speed-improvement over naive brute-force attacks, was demonstrated, showing its potential to allow for computable brute-force attempts.
Future work should explore further experimentation with different pre- and post-processing techniques, as well as feature combinations to improve clustering accuracy. Additionally, techniques like Metropolis-Hastings for probabilistically improving clusters could be tested, as seen in the Open Source KeyTap2 project~\cite{Keytap2_Github}. The impact of adding complexity to typing (e.g., special characters, uppercase letters, and backspaces) should also be explored to assess the attack’s feasibility under more realistic conditions. Furthermore, the data in this work shows that participants were not equally susceptible to the attack and future work should target specific reasons for why this may be, such as typing style, microphone quality and the used keyboard.

With recommendations from agencies like the NCSC advising three-word passphrases, the attack in this work presents a potential risk, underscoring the need for improved passphrase security through varied delimiters, special characters, and increased randomisation. 

\printbibliography

@misc{sabra_zoom_2020,
	title = {Zoom on the {Keystrokes}: {Exploiting} {Video} {Calls} for {Keystroke} {Inference} {Attacks}},
	shorttitle = {Zoom on the {Keystrokes}},
	url = {http://arxiv.org/abs/2010.12078},
	language = {en},
	urldate = {2024-05-27},
	publisher = {arXiv},
	author = {Sabra, Mohd and Maiti, Anindya and Jadliwala, Murtuza},
	month = oct,
	year = {2020},
	note = {arXiv:2010.12078 [cs]},
}

@online{NCSC_password,  
    author= {{National Cyber Security Centre}},  
    title= {{Top tips for staying secure online}},
    url={https://www.ncsc.gov.uk/collection/top-tips-for-staying-secure-online/three-random-words},
    journal={NCSC}, urldate= {2024-08-29},
    year={2021},
    month={12},
    day={21},
}

@online{Diceware,author={Douglas Muth},title={{Diceware Password Generator}}, url={https://diceware.dmuth.org/}, journal={Diceware},urldate   = {2025-03-16}, }

@online{EFF_dware, author={{Electronic Frontier Foundation}}, title={{EFF Dice-Generated Passphrases
}}, url={https://www.eff.org/de/dice}, journal={Electronic Frontier Foundation}, year={2023}, month={1}, urldate   = {2025-03-16},}

@inproceedings{barisani2009sniffing,
  title={Sniffing Keystrokes With Lasers and Voltmeters},
  author={Barisani, A. and Bianco, D.},
  booktitle={Proceedings of Black Hat USA},
  year={2009},
  organization={Black Hat},
  url={https://www.blackhat.com/presentations/bh-usa-09/BARISANI/BHUSA09-Barisani-Keystrokes-PAPER.pdf},
}

@inproceedings{vuagnoux2009compromising,
  title={Compromising electromagnetic emanations of wired and wireless keyboards.},
  author={Vuagnoux, Martin and Pasini, Sylvain},
  booktitle={USENIX security symposium},
  volume={8},
  pages={1--16},
  year={2009}
}

@ARTICLE{WirelessTraining-FreeKeystrokeInferenceAttackandDefense,
  author={Yang, Edwin and Fang, Song and Markwood, Ian and Liu, Yao and Zhao, Shangqing and Lu, Zhuo and Zhu, Haojin},
  journal={IEEE/ACM Transactions on Networking}, 
  title={Wireless Training-Free Keystroke Inference Attack and Defense}, 
  year={2022},
  volume={30},
  number={4},
  pages={1733-1748},
  doi={10.1109/TNET.2022.3147721}}

@inproceedings{zhu2014context,
  title={Context-free attacks using keyboard acoustic emanations},
  author={Zhu, Tong and Ma, Qiang and Zhang, Shanfeng and Liu, Yunhao},
  booktitle={Proceedings of the 2014 ACM SIGSAC conference on computer and communications security},
  pages={453--464},
  year={2014}
}

@article{zhuang_keyboard_2009,
	title = {Keyboard acoustic emanations revisited},
	volume = {13},
	issn = {1094-9224},
	url = {https://doi.org/10.1145/1609956.1609959},
	doi = {10.1145/1609956.1609959},
	number = {1},
	urldate = {2024-07-24},
	journal = {ACM Trans. Inf. Syst. Secur.},
	author = {Zhuang, Li and Zhou, Feng and Tygar, J. D.},
	month = nov,
	year = {2009},
	pages = {3:1--3:26},
}

@inproceedings{asonov_keyboard_2004,
	title = {Keyboard acoustic emanations},
	url = {https://ieeexplore.ieee.org/abstract/document/1301311},
	doi = {10.1109/SECPRI.2004.1301311},
	urldate = {2024-07-20},
	booktitle = {{IEEE} {Symposium} on {Security} and {Privacy}, 2004. {Proceedings}. 2004},
	author = {Asonov, D. and Agrawal, R.},
	month = may,
	year = {2004},
	note = {ISSN: 1081-6011},
	pages = {3--11},
}

@inproceedings{Berger_dictionary_acoustic_emantions,
author = {Berger, Yigael and Wool, Avishai and Yeredor, Arie},
title = {Dictionary attacks using keyboard acoustic emanations},
year = {2006},
isbn = {1595935185},
publisher = {Association for Computing Machinery},
address = {New York, NY, USA},
url = {https://doi.org/10.1145/1180405.1180436},
doi = {10.1145/1180405.1180436},
booktitle = {Proceedings of the 13th ACM Conference on Computer and Communications Security},
pages = {245–254},
numpages = {10},
location = {Alexandria, Virginia, USA},
series = {CCS '06}
}

@online{Keytap2_Github,
  author       = {Georgi Gerganov},
  title        = {Keytap2 - Acoustic Keyboard Eavesdropping Based on Language N-gram Frequencies},
  year         = {2020},
  month        = {12},
  url          = {https://github.com/ggerganov/kbd-audio/discussions/31},
  organization = {GitHub Discussions},
  urldate = {2025-03-13}
}

@inproceedings{liu_snooping_2015,
	address = {Paris France},
	title = {Snooping {Keystrokes} with mm-level {Audio} {Ranging} on a {Single} {Phone}},
	isbn = {978-1-4503-3619-2},
	url = {https://dl.acm.org/doi/10.1145/2789168.2790122},
	doi = {10.1145/2789168.2790122},
	language = {en},
	urldate = {2024-05-22},
	booktitle = {Proceedings of the 21st {Annual} {International} {Conference} on {Mobile} {Computing} and {Networking}},
	publisher = {ACM},
	author = {Liu, Jian and Wang, Yan and Kar, Gorkem and Chen, Yingying and Yang, Jie and Gruteser, Marco},
	month = sep,
	year = {2015},
	pages = {142--154},
}

@inproceedings{murali2018keyboard, title={Keyboard side channel attacks on smartphones using sensor fusion}, author={Murali, Nithin and Appaiah, Kumar}, booktitle={2018 IEEE Global Communications Conference (GLOBECOM)}, pages={206--212}, year={2018}, organization={IEEE} }

@inproceedings{halevi2012closer,
  title={A closer look at keyboard acoustic emanations: random passwords, typing styles and decoding techniques},
  author={Halevi, Tzipora and Saxena, Nitesh},
  booktitle={Proceedings of the 7th ACM Symposium on Information, Computer and Communications Security},
  pages={89--90},
  year={2012}
}

@inproceedings{10.1145/3359789.3359816,
author = {Slater, David and Novotney, Scott and Moore, Jessica and Morgan, Sean and Tenaglia, Scott},
title = {Robust keystroke transcription from the acoustic side-channel},
year = {2019},
isbn = {9781450376280},
publisher = {Association for Computing Machinery},
address = {New York, NY, USA},
url = {https://doi.org/10.1145/3359789.3359816},
doi = {10.1145/3359789.3359816},
booktitle = {Proceedings of the 35th Annual Computer Security Applications Conference},
pages = {776–787},
numpages = {12},
location = {San Juan, Puerto Rico, USA},
series = {ACSAC '19}
}

@inproceedings{harrison2023practical,
  title={A practical deep learning-based acoustic side channel attack on keyboards},
  author={Harrison, Joshua and Toreini, Ehsan and Mehrnezhad, Maryam},
  booktitle={2023 IEEE European Symposium on Security and Privacy Workshops (EuroS\&PW)},
  pages={270--280},
  year={2023},
  organization={IEEE}
}

@article{pleva_acoustical_2015,
	title = {Acoustical {User} {Identification} {Based} on {MFCC} {Analysis} of {Keystrokes}},
	volume = {13},
	copyright = {The authors publishing with this Journal agree to the following terms and conditions:   The authors retain copyright and grant the Journal right of the first publication with the work simultaneously licensed under a  Creative Commons Attribution License  that allows others to share the work with an acknowledgement of the work's authorship and initial publication in this Journal.    The authors are able to enter into separate, additional contractual arrangements for the non-exclusive distribution of the Journal's published version of the work (e.g., to post it to an institutional repository or publish it in a book), with an acknowledgement of its initial publication in this Journal.    The authors are permitted and encouraged to post their work online (e.g., in institutional repositories or on their websites) prior to and during the submission process, as it can lead to productive exchanges, as well as earlier and greater citation of the published work (See  The Effect of Open Access ).    From the first issue of the 2021 the Creative Commons Attribution License (in our case it´s BY-CC) under which are publishing articles in Advances in Electrical and Electronic Engineering also appear in the full text of all published articles.},
	issn = {1804-3119},
	url = {http://advances.utc.sk/index.php/AEEE/article/view/1466},
	doi = {10.15598/aeee.v13i4.1466},
	language = {en},
	number = {4},
	urldate = {2024-07-20},
	journal = {Advances in Electrical and Electronic Engineering},
	author = {Pleva, Matus and Kiktova, Eva and Juhar, Jozef and Bours, Patrick},
	month = nov,
	year = {2015},
	note = {Number: 4},
	pages = {309--313},
}

@inproceedings{Bonneau12,
  author = {Bonneau, Joseph},
  booktitle = {IEEE Symposium on Security and Privacy},
  ee = {https://doi.ieeecomputersociety.org/10.1109/SP.2012.49},
  isbn = {978-0-7695-4681-0},
  pages = {538-552},
  publisher = {IEEE Computer Society},
  title = {The Science of Guessing: Analyzing an Anonymized Corpus of 70 Million Passwords.},
  url = {http://dblp.uni-trier.de/db/conf/sp/sp2012.html#Bonneau12},
  year = 2012
}

@inproceedings{owusu_accessory_2012,
	address = {New York, NY, USA},
	series = {{HotMobile} '12},
	title = {{ACCessory}: password inference using accelerometers on smartphones},
	isbn = {978-1-4503-1207-3},
	shorttitle = {{ACCessory}},
	url = {https://doi.org/10.1145/2162081.2162095},
	doi = {10.1145/2162081.2162095},
	urldate = {2024-07-20},
	booktitle = {Proceedings of the {Twelfth} {Workshop} on {Mobile} {Computing} {Systems} \& {Applications}},
	publisher = {Association for Computing Machinery},
	author = {Owusu, Emmanuel and Han, Jun and Das, Sauvik and Perrig, Adrian and Zhang, Joy},
	month = feb,
	year = {2012},
	pages = {1--6},
}

@inproceedings{vibration_inference,
author = {Marquardt, Philip and Verma, Arunabh and Carter, Henry and Traynor, Patrick},
year = {2011},
month = {10},
pages = {551-562},
title = {(sp)iPhone: Decoding vibrations from nearby keyboards using mobile phone accelerometers},
journal = {Proceedings of the ACM Conference on Computer and Communications Security},
doi = {10.1145/2046707.2046771}
}

@article{passwd_sec_a_case_history,
author = {Morris, Robert and Thompson, Ken},
title = {Password security: a case history},
year = {1979},
issue_date = {Nov. 1979},
publisher = {Association for Computing Machinery},
address = {New York, NY, USA},
volume = {22},
number = {11},
issn = {0001-0782},
url = {https://doi.org/10.1145/359168.359172},
doi = {10.1145/359168.359172},
journal = {Commun. ACM},
month = {11},
pages = {594–597},
numpages = {4}
}

@book{wikey_2015,
	title = {Keystroke {Recognition} {Using} {WiFi} {Signals}},
	author = {Ali, Kamran and Liu, Alex and Wang, Wei and Shahzad, Muhammad},
	publisher      = {IEEE},
	month = sep,
	year = {2015},
	doi = {10.1145/2789168.2790109},
}

@inproceedings{dhakal2018observations,
author = {Dhakal, Vivek and Feit, Anna and Kristensson, Per Ola and Oulasvirta, Antti},
booktitle = {Proceedings of the 2018 CHI Conference on Human Factors in Computing Systems (CHI '18)},
title = {Observations on Typing from 136 Million Keystrokes},
year = {2018},
publisher = {ACM},
doi = {https://doi.org/10.1145/3173574.3174220}
}

@inproceedings{how_we_type,
author = {Feit, Anna Maria and Weir, Daryl and Oulasvirta, Antti},
title = {How We Type: Movement Strategies and Performance in Everyday Typing},
year = {2016},
isbn = {9781450333627},
publisher = {Association for Computing Machinery},
address = {New York, NY, USA},
url = {https://doi.org/10.1145/2858036.2858233},
doi = {10.1145/2858036.2858233},
booktitle = {Proceedings of the 2016 CHI Conference on Human Factors in Computing Systems},
pages = {4262–4273},
numpages = {12},
location = {San Jose, California, USA},
series = {CHI '16}
}

\onecolumn
\appendix
\section{Appendix} \label{sec:app-a}
\subsection{Generated Passphrases}\label{sec:passphrases}

The following passphrases were used in the experiments:
\vspace{0.4cm}

\begin{minipage}{0.45\textwidth}
    \footnotesize
    \setcounter{enumi}{-1}
        \begin{enumerate}
        \item peroxide hacking arena
        \item goldfish augmented yoyo
        \item nugget iguana nylon
        \item finalist caviar cufflink
        \item ipad decal uptown
        \item lukewarm pedometer litter wreckage
        \item juggle gibberish hacking luxurious
        \item unmarked vaseline aluminum jasmine
        \item poison amendment sizable angelfish
        \item taco ferret circle deliverer
        \item velcro jelly duplex magazine silicon
        \item hefty frosting acid zookeeper patio
        \item daughter pyramid onyx pogo palm
        \item cahoots arena cement statue mutation
        \item blade banana awhile elsewhere tadpole
  
    \end{enumerate}
\end{minipage}%
\hfill
\begin{minipage}{0.45\textwidth}
    \begin{enumerate}
        \footnotesize
        \setcounter{enumi}{15}
        \item oxygen remote diffuser engine lettuce acid
        \item oncoming feline glucose sushi abdomen judiciary
        \item nullify scarf deepness modify euphemism grumbling
        \item apple unnoticed bullfrog datebook vicinity glove
        \item unhinge zodiac movie tadpole tapestry waffle
        \item habitat gullible jingling mule envoy device erratic
        \item licorice breath thumb navigate saddlebag yahoo voucher
        \item festival yearbook fountain underwear nastiness dedicate licorice
        \item scooter urchin albatross sneezing itunes gumdrop cubical
        \item bagpipe earlobe aerosol aliens ivory clubhouse pantyhose
        \item couch crawfish mundane goggles rupture florist rancidity degree
        \item hefty tree riverboat sculpture junkyard awhile isotope unveiled
        \item sled dyslexia jelly clergyman fruit family blade rancidity
        \item payphone rupture awoke virus tuesday upbeat knapsack amnesty
        \item afloat ardently fox emission exquisite dagger jersey lubricant
    \end{enumerate}
\end{minipage}

\subsection{Differing amounts of clusters for demodulation}\label{app:single_cluster_recovery}

\begin{figure}[!ht]
    \centering
    \begin{minipage}{0.5\textwidth}
        \centering
        \includegraphics[width=1\textwidth]{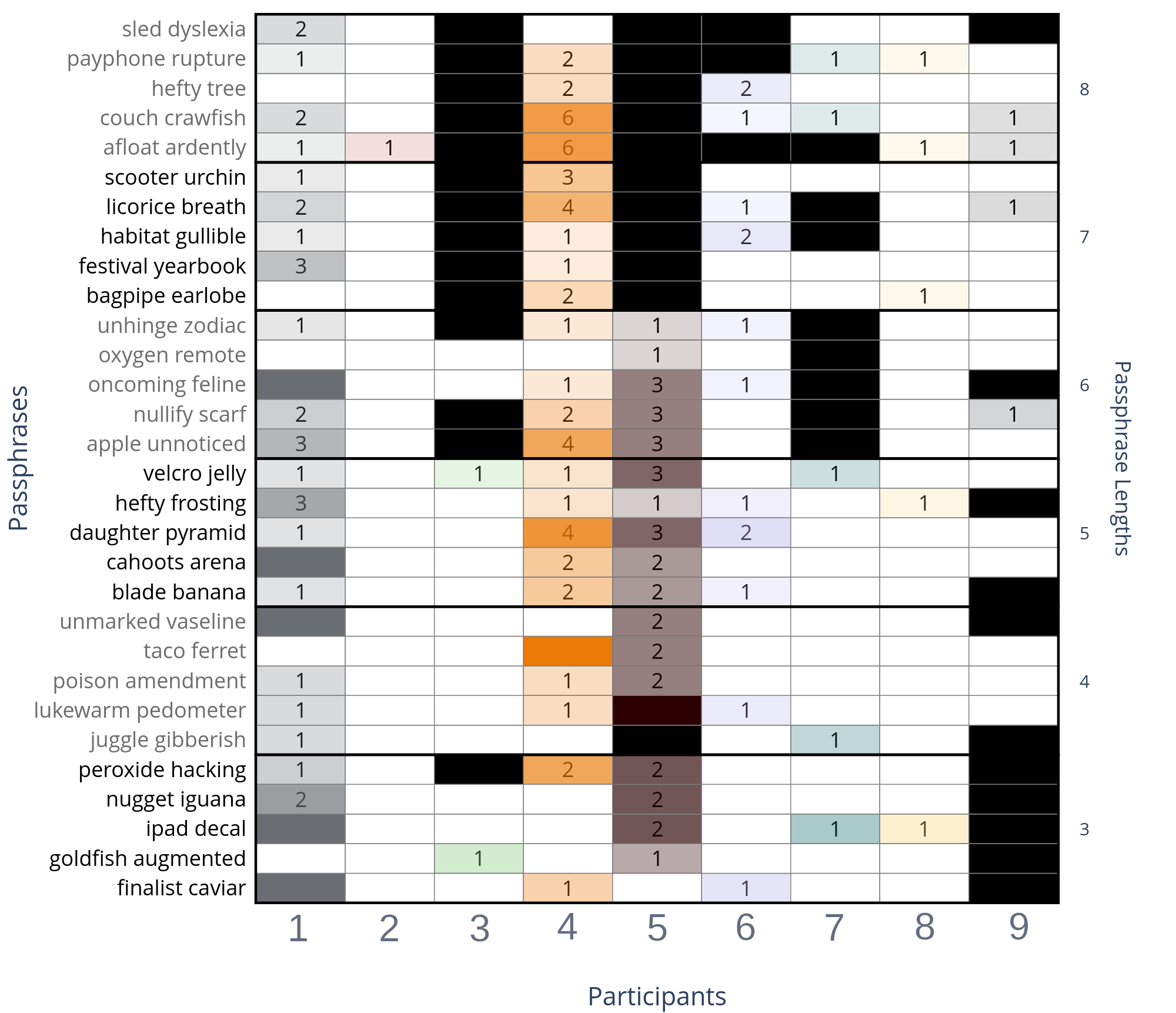}
        \caption{Recovery results using one set of clusters from the best model.}
        \label{fig:recovery_results_rectangles}
    \end{minipage}%
    \hfill
    \begin{minipage}{0.5\textwidth}
        \centering
        \includegraphics[width=1\textwidth]{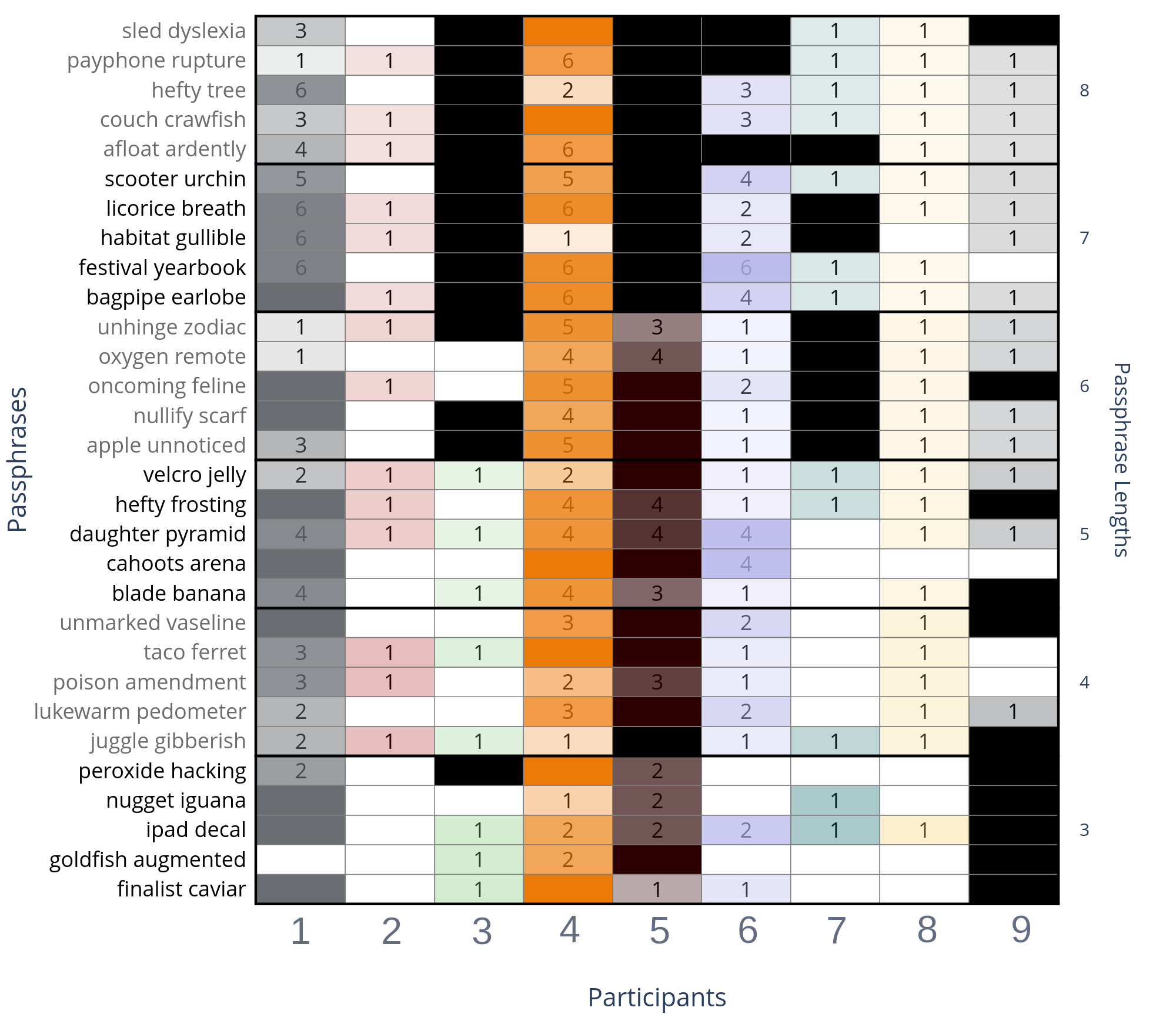}
        \caption{Recovery results using 50 clusters.}
        \label{fig:recovery_results_rectangles_fifty_clusterings}
    \end{minipage}
\end{figure}

\begin{figure}[!ht]
    \centering
    \includegraphics[width=0.85\textwidth]{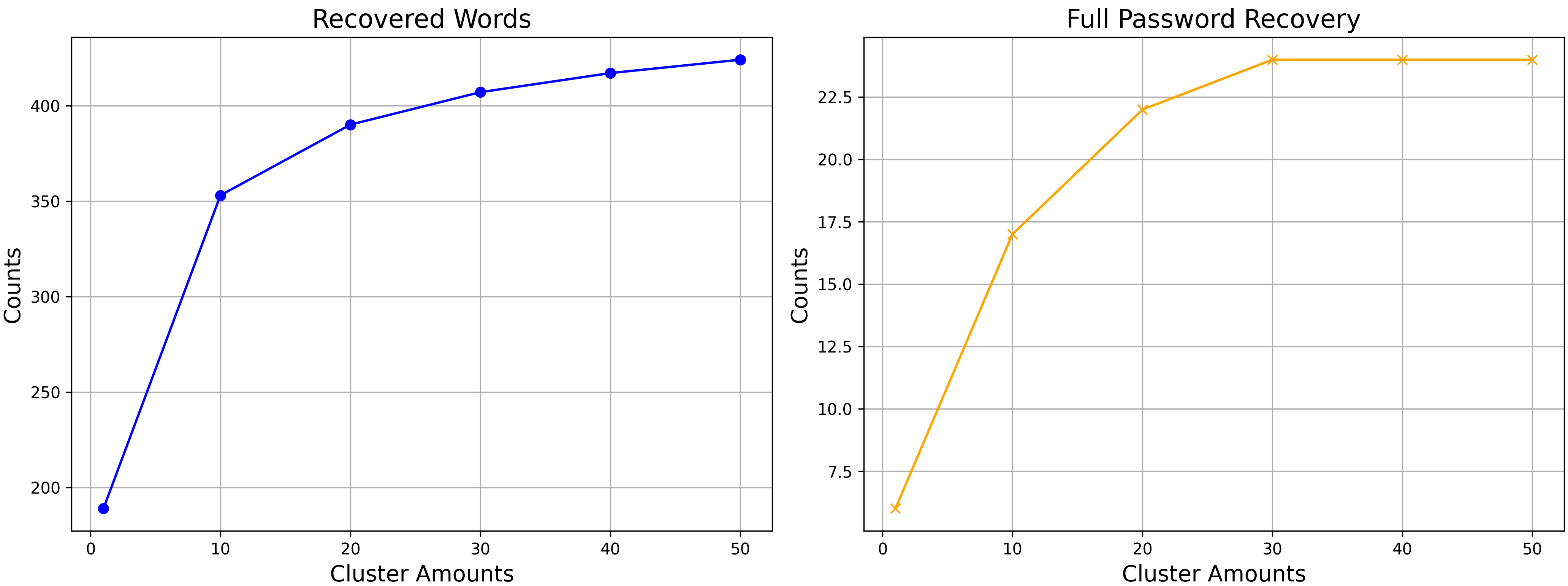}
    \caption{Recovery as a function of clusters.}
    \label{fig:recovery_results_recovery_relation_to_clusters}
\end{figure}

\end{document}